\documentclass[superscriptaddress,
 reprint,
 amssymb, amsmath,
aps,prl 
]{revtex4-1}
\usepackage{docs}%
\usepackage{dcolumn}
\usepackage{amsmath}
\usepackage{amsfonts}
\usepackage{amssymb}
\usepackage{graphicx,color}
\usepackage{mathtools}
\usepackage{bm}
\usepackage{algorithm}
\usepackage{algpseudocode}
\usepackage[breaklinks=true]{hyperref}
\hypersetup{
  colorlinks   = true, %Colours links instead of ugly boxes
  urlcolor     = blue, %Colour for external hyperlinks
  linkcolor    = blue, %Colour of internal links
  citecolor   = blue %Colour of citations
}

\graphicspath{{figures/}}

\begin{document}

\title{Chaos and the Flow Capture Problem: Polluting is Easy, Cleaning is Hard}

\author{Lachlan~D. Smith}
 \email{lachlan.smith@northwestern.edu}
 \affiliation{\mbox{Department of Chemical and Biological Engineering,
   Northwestern University, Evanston, IL 60208, USA}}
\author{Guy Metcalfe}
 \email{gmetcalfe@swin.edu.au}
 \affiliation{Department of Mechanical and Product Design
  Engineering, Swinburne University of Technology, Hawthorn, VIC 3122,
  Australia}
\author{Julio~M. Ottino}
 \email{jm-ottino@northwestern.edu}
 \affiliation{\mbox{Department of Chemical and Biological Engineering,
   Northwestern University, Evanston, IL 60208, USA}}
 \affiliation{\mbox{Department of Mechanical Engineering, Northwestern
   University, Evanston, IL 60208, USA}}
 \affiliation{\mbox{The Northwestern Institute on Complex Systems (NICO),
  Northwestern University, Evanston, IL 60208, USA}}

\date{\today}

\begin{abstract}
  Cleaning pollution from a heterogeneous flow environment is far from simple.  We
  consider the flow capture problem, which has flows and sinks in a
  heterogeneous environment, and investigate the problem of positioning pollutant capture units.  We show that arrays of capture units
  carry a high risk of failure without accounting for environmental
  heterogeneity and chaos in their placement, design, and operation.  Our idealized 2-dimensional models reveal salient
  features of the problem.  Maximum capture efficiency depends on the
  required capture rate: long term efficiency decreases as the number
  of capture units increases, whereas short term efficiency
  increases. If efficiency is important, the capture process should
  begin as early as feasible.  Knowledge of transport controlling
  flow structures offers predictability for unit placement.  We demonstrate two heuristic approaches to near-optimally position capture units.
\end{abstract}

\maketitle

\section{Introduction}

Consider the problem of positioning units in a heterogeneous flow
environment to capture something moved by the flow through the
environment.  We label this the \emph{flow capture problem}, which has
many applications.  For example, in the ocean mussels massively
congregate on the pilings of wind farms.  An unintended consequence of
this build-up is that a large portion of phytoplankton is filtered out
of the surrounding water, leading to substantial change in local ocean
ecology \cite{Slavik_pelagic_2017}.  Another example is removal of a
pollutant from an environment, e.g.\ microplastics that are
accumulating in oceans and waterways \cite{Andrady2011microplastics,
  Cozar2014microplastics}, or capturing CO$_2$ directly from the
atmosphere in an effort to limit global temperature rise
\cite{Socolow2011direct, IPCC_2014, Smith_limits_2015, Keith2018}.  The central question in all these examples is:
Where should capturing units be placed in order to meet specified
objectives, e.g.\ minimizing the impact of mussel congregation, or
maximizing pollutant removal?  All these cases are specific examples
of the flow capture problem, problems involving flows and sinks.  The
most challenging capture problems involve complex flows.

The flow capture problem can be cast in two ways.  The first maximizes
the total amount of captured material after some fixed length of time.
This setting resembles facility location optimization problems
\cite{Bansal2017MCLP}, for example sensor placement in water or
ventilation distribution systems to detect a maliciously injected
contaminant \cite{Krause2008} or in cell-phone contact networks to
detect a virus \cite{Lee_detection_2015}, problems in principle
solvable using large-scale mixed integer linear programming.  But
there are several things that make the general flow capture problem
more challenging.  The shapes of the captured regions can in general
be very complex, possibly fractal, and need not be convex or simply
connected (they have holes).  Moreover, finding the region that is
captured by a unit is difficult.  The second type of flow capture
problem optimizes unit placement to minimize the time to reach a
critical value of capture (maximizing the capture rate).  This problem
accounts for time-dependent capture, i.e.\/ the capture zone for a
unit increases as a function of time, and cannot easily be cast as a
facility optimization problem.  The capture rate can also be
understood by considering capture units as the leaks in leaking
chaotic systems \cite{Altmann2013leaking}.  Over long times, the
capture rate decays exponentially at a predictable rate.  However,
over short times the capture rate is much less predictable and is
highly sensitive to a unit's location.

We consider the placement of perfect capture units in three scenarios and estimate how effective cleaning will be in each case.
\begin{enumerate}
\item Steady homogeneous flow.  No chaos; just uniform, unidirectional
  flow.
\item Complex heterogeneous flow, but we assume no knowledge of the
  flow, i.e.\/ placement choice is blind to local flow heterogeneity;
\item Complex heterogeneous flow, but we use knowledge of the flow to
  inform placement.
\end{enumerate}
Define $P(t)$ as the fraction of pollutant concentration in the domain
at time $t$, and assume an initially uniform concentration:
$P(t = 0) = 1$.  Define $P_c$ as the target value of $P$ that we want
to reduce to.  The effectiveness of $N$ removal units is determined by
whether or not they can reduce $P$ to less than $P_c$ in finite time:
$P \le P_c$ defines success, while $P > P_c$ for all $t$ is failure.

\section{Uniform flow}

First consider steady flow on a
doubly periodic square domain of side length $h$.  For each fixed
number of units $N$, each unit interacts with only a small, fixed
swathe of the domain: $P$ decreases linearly from $1$, levels
out to some value, and stays there for all time.  Once a unit's swathe
is exhausted of pollutant, no new pollutant ever comes to that unit.
Each unit captures an area equal to $\delta h$, where $\delta$ is the
unit characteristic capture size.  Therefore, the equilibrium value of
$P$ is $1-N\delta /h$, assuming the units do not overlap.  As $N$
grows, the equilibrium value of $P$ decreases.  For
$N \geq N_c = \frac{h}{\delta}(1-P_c)$, $P$ will be lower than $P_c$,
and cleaning will be successful.  We can define a long-time measure of
effectiveness for the steady flow as
\begin{equation}
\label{eq:effectivness_steady}  
  E(N) =
  \begin{cases}
    0 & \quad \mathrm{for} \quad N \quad \mathrm{such \: that} \quad P > P_c \\
    1 & \quad \mathrm{for} \quad N \quad \mathrm{such \: that} \quad P \le P_c,
  \end{cases}
\end{equation}
and $E(N)=0$ for $N<N_c$, $E(N)=1$ for $N\geq N_c$.

\section{Effectiveness of random placement in complex flows}

\begin{figure}[t]
\centering
\includegraphics[width=0.45\textwidth]{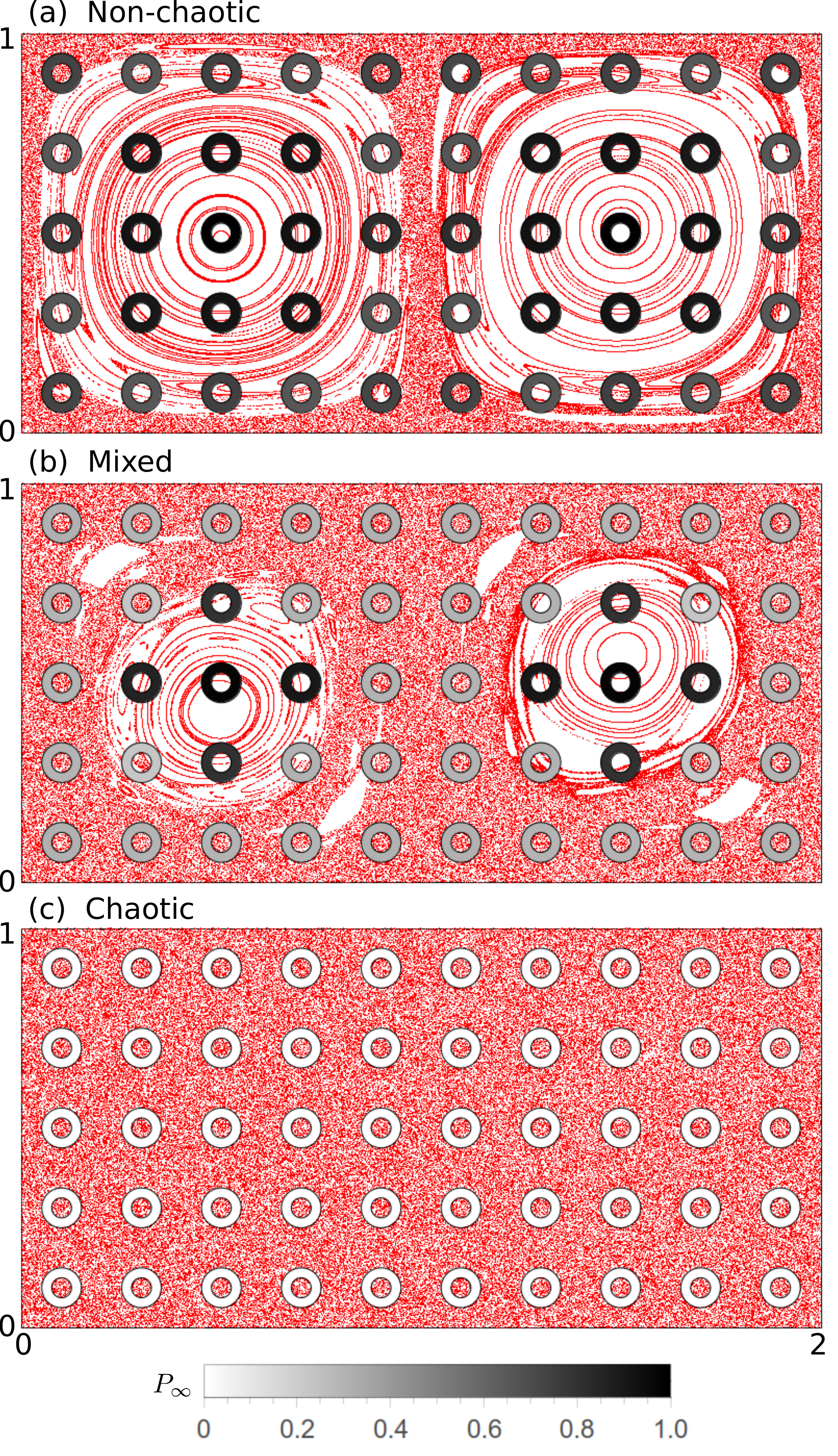}
\caption{Poincar\'{e} sections (red) generated by the double gyre flow eq.~(\ref{eq:double_gyre}) for (a)~$\epsilon=0.01$, which is mostly non-chaotic, (b)~$\epsilon=0.05$, which has approximately equal areas of chaotic and non-chaotic regions, and (c)~$\epsilon=0.25$, which is fully chaotic.  The circles, of diameter $\delta=0.1$, are possible sites for
  capture units, which are colored according to $P_\infty$ for each
  unit run in isolation, i.e.\/ $N=1$.  No pollutant removal means $P_\infty=1$, shown as black, and total removal means $P_\infty=0$, shown as white, which is achieved by all locations in (c).}
\label{fig:colored_machine_locations}
\end{figure}

What about complex two-dimensional flows, with non-mixing ``islands'' interspersed amongst a chaotic ``sea''?  We first pretend we know nothing about
the underlying flow, but we can measure $P(t)$.  We consider the double gyre flow \cite{Shadden2005}, a model for two gyre systems observed in geophysical flows, in three representative cases. The time-periodic velocity is given by
\begin{align}
v_x(x,y,t) &=  -\frac{\pi}{2}\sin\left(\pi f(x,t)\right) \cos\left( \pi y\right) \nonumber \\
v_y(x,y,t) &=  \frac{\pi}{2} \cos\left(\pi f(x,t) \right) \sin\left(\pi y\right) \frac{\partial f}{\partial x}\left( x,t \right) \label{eq:double_gyre} \\
f(x,t) &= \epsilon \sin(2\pi t) x^2 + \left(1-2\epsilon \sin(2\pi t) \right) x,  \nonumber 
\end{align}
with time-period equal to 1.  For $\epsilon=0.01$, the flow is mostly
non-chaotic, as shown by the Poincar\'{e} section \footnote{The
  positions of a number of tracer particles are marked at the end of
  each flow period.} (red) in
Fig.~\ref{fig:colored_machine_locations}(a), which is dominated by two
non-mixing islands. We label this case ``non-chaotic.'' For
$\epsilon=0.05$, the chaotic sea has approximately the same area as
the non-chaotic islands, as shown in
Fig.~\ref{fig:colored_machine_locations}(b), and we label this case
``mixed.'' Lastly, for $\epsilon=0.25$, the flow is fully chaotic,
evidenced by the lack of islands in
Fig.~\ref{fig:colored_machine_locations}(c). We label this case
``chaotic.'' In each figure of
Fig.~\ref{fig:colored_machine_locations} the circles are $J = 50$
capture units (sinks) of equal diameter $\delta=0.1$ (the shading is
explained later).  Here we only consider combinations of these 50
locations; however, the complete flow capture problem allows each site
to be located \emph{anywhere} in the domain.  Any pollutant that
reaches a sink is immediately removed from the domain.  For each value
of $N$ and each flow case, we measure $P(t)$ for all combinations of
$N$ units from the grid of possible sites shown in
Fig.~\ref{fig:colored_machine_locations}, to obtain ensemble
statistics.  Note the explosion of possible site combinations.  Even
for our simple example with just $50$ possible sites, for $N=5$ there
are over 2 million possible combinations. Moreover, the reality is
that we would not be able to build $N$ sites, only to tear them down
and build them afresh until we discovered the optimum placement.  We
get only one pass through the distribution of possible outcomes.

\begin{figure}[tb]
\centering
\includegraphics[width=0.45\textwidth]{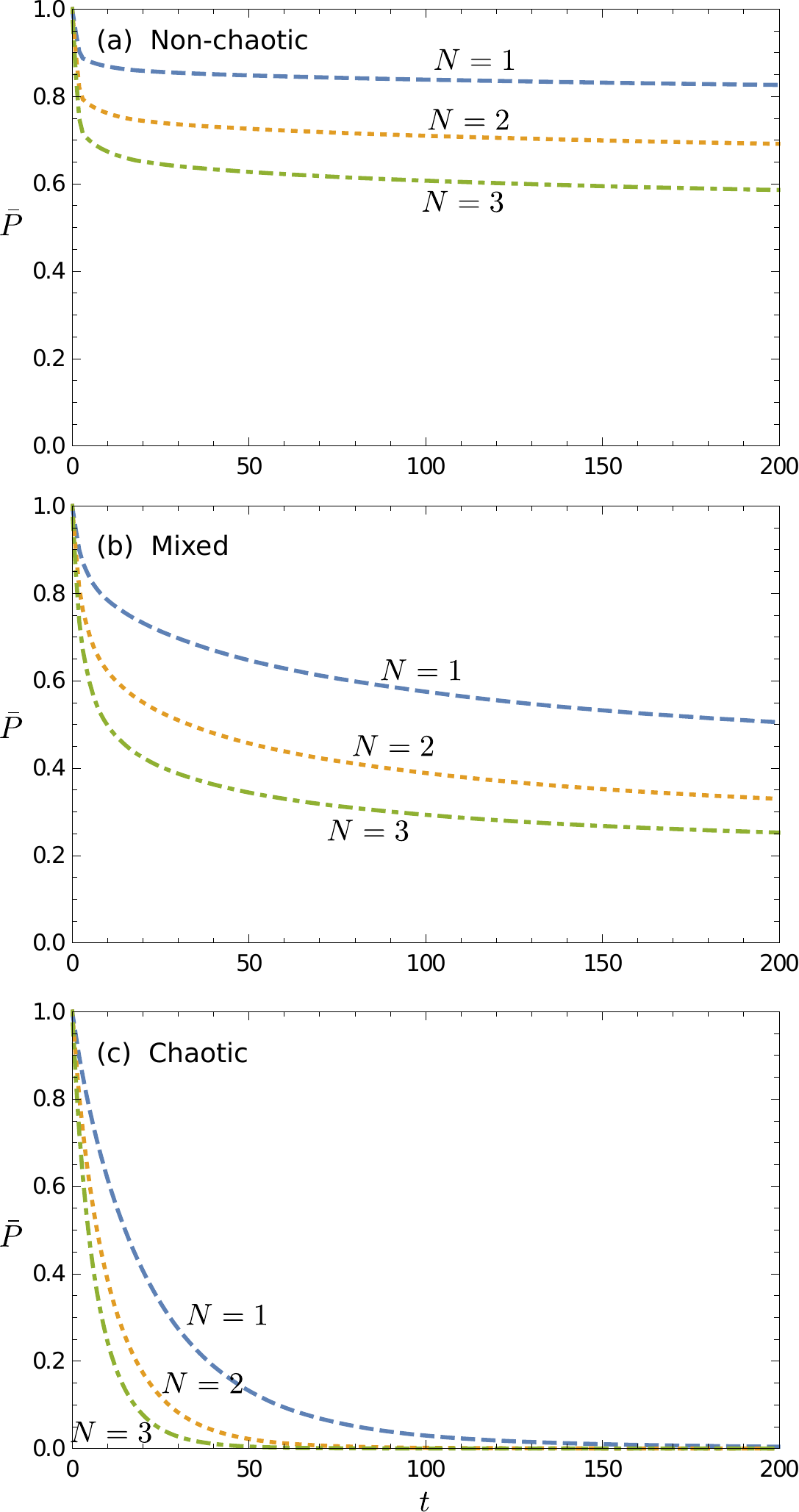}
\caption{Averages of $P(t)$ over all combinations of $N$ sites from
  the grid in Fig.~\protect\ref{fig:colored_machine_locations}, for
  $N=1,2$, and $3$, and the three cases shown in Fig.~\ref{fig:colored_machine_locations}, (a)~non-chaotic, (b)~mixed, and (c)~chaotic.  As we choose $N$ sites from $J=50$ possible sites,
  the number of possible combinations grows as the binomial
  coefficient, rising rapidly from 50 combinations for $N = 1$ to
  over 2 million for $N = 5$.}
\label{fig:mean_concentration-vs-time}
\end{figure}

\begin{figure}[tp]
\centering
\includegraphics[width=0.45\textwidth]{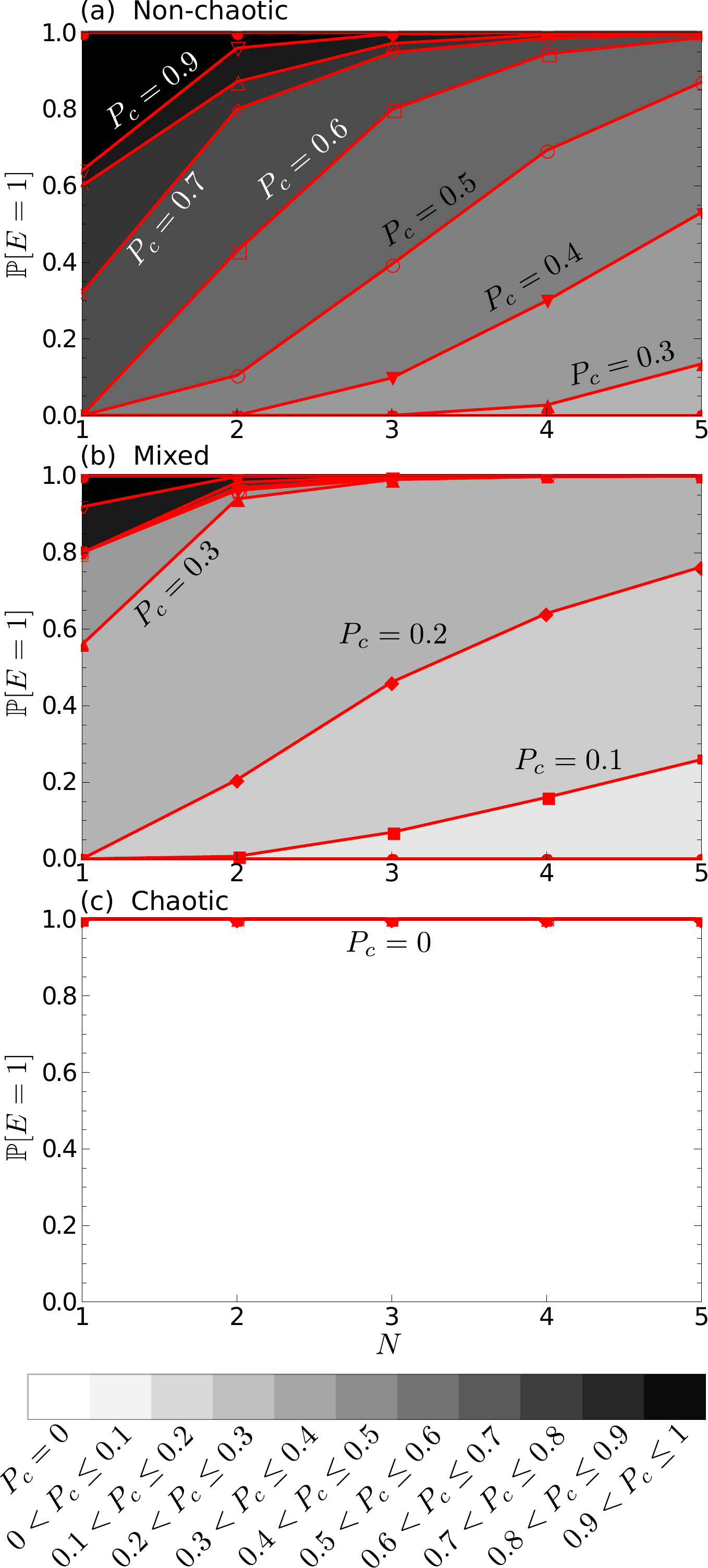}
\caption{Probability of success of a random combination of $N$ sites,
  i.e.\/ probability that the combination will achieve $E=1$ from
  Eq.~\protect\ref{eq:effectivness_chaotic}, for each of the three cases, (a)~non-chaotic, (b)~mixed, and (c)~chaotic.  Each red curve corresponds to a different value of $P_c$, ranging from $0$ to $1$ in increments of $0.1$.}
\label{fig:probability_of_effectiveness}
\end{figure}

Fig.~\ref{fig:mean_concentration-vs-time} shows the ensemble averages
of $P(t)$ for the three cases; however, the averages hide as much as
they reveal.  As $N$ increases, more pollutant is removed on average,
but for each $N$, the long-time value of $P$, $P_\infty$ \footnote{We
  use $P(1000)$ to approximate $P_\infty$, i.e.\ 1,000 flow periods.},
for each combination of sites can diverge strongly from the average.
Some site combinations achieve values of $P_\infty$ much lower than
$P_c$, or reach $P_c$ very quickly, while other combinations never
reach $P_c$.  Like eq.~(\ref{eq:effectivness_steady}), we define a
long-time effectiveness for each combination of $N$ sites,
$(m_1, \dots, m_N)$, as
\begin{equation}
\label{eq:effectivness_chaotic}  
  E(m_1,\dots,m_N) =
  \begin{cases}
    0 & \quad \mathrm{if} \quad P_\infty > P_c \\
    1 & \quad \mathrm{if} \quad P_\infty \le P_c
  \end{cases}
\end{equation}
and plot the probability that a random combination of $N$ sites will
be effective, $\mathbb{P}[E=1]$, in
Fig.~\ref{fig:probability_of_effectiveness}.  For the non-chaotic case [Fig.~\ref{fig:probability_of_effectiveness}(a)], not surprisingly, when
$N$ and $P_c$ are small, few combinations of sites will be effective, while for
large $N$ most combinations will be effective.  For instance, for $N=1$, no individual site is capable of reaching $P_c\leq 0.6$, whereas almost all combinations of $5$ sites achieve $P_c=0.6$. Note that for $N=5$, no combination is able to achieve $P_c\leq 0.2$.  

In stark contrast, in the fully chaotic case [Fig.~\ref{fig:probability_of_effectiveness}(c)], \emph{every} individual site cleans the entire domain, meaning $\mathbb{P}[E=1]=1$ for every value of $P_c$. This is because dynamics in the chaotic region are ergodic, i.e.\ trajectories of tracer particles densely fill the chaotic region.  Therefore, every pollutant particle will eventually be captured, regardless of the position of the capture unit.  This shows that the biggest challenge to long-term removal is the presence of non-chaotic regions.  However, for shorter time-frames, the capture sites in the chaotic case do not clean the entire domain, and some sites capture more than others.  This means optimal siting is important if cleaning must occur rapidly.

The mixed case [Fig.~\ref{fig:probability_of_effectiveness}(b)] falls in between the two extremes.  Single units ($N=1$) are not able to clean the entire domain, but they are able to achieve smaller values of $P_c$ than the non-chaotic case.  Note that even when $N=5$, no combination can clean the entire domain.  This is linked to the size of the two non-chaotic islands, which have diameter slightly greater than five times the diameter of the capture unit.  

For the non-chaotic and mixed cases, failure occurs even though our model capture units are
assumed perfect at cleaning the fluid they receive.  Failure is
entirely a function of the heterogeneity of the environment, in particular non-chaotic islands.  The
clear message is that for all but fully chaotic flows, the more ambitious the pollution reduction
target, the more dominant becomes consideration of site location.

\section{Informed placement based on flow characteristics}

\subsection{Maximum capture efficiency}

\begin{figure}[tp]
\centering
\includegraphics[width=0.45\textwidth]{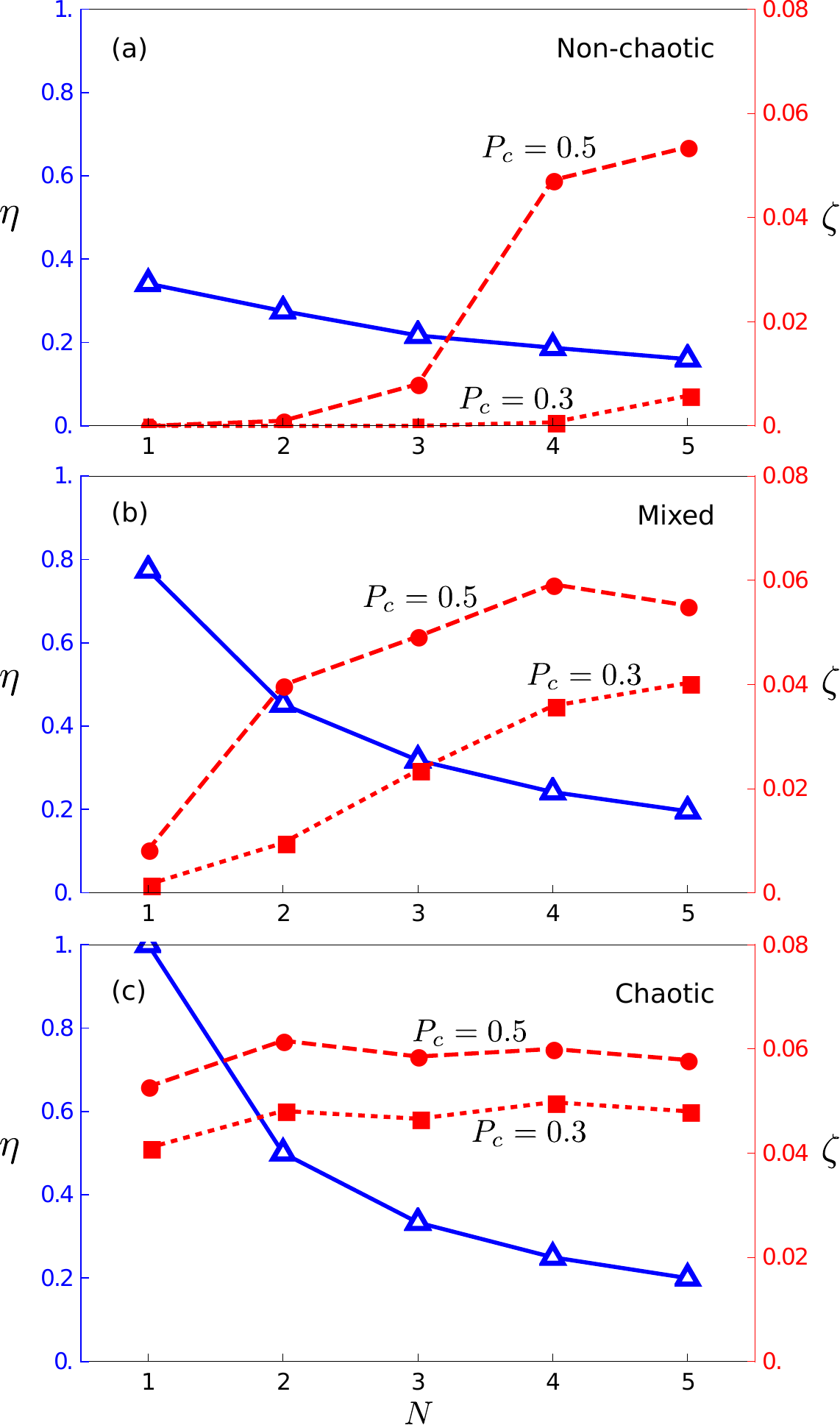}
\caption{Efficiency measures for $N$ capturing units in each of the three cases, (a)~non-chaotic, (b)~mixed, and (c)~chaotic.  Maximum $\eta$
  efficiency of $N$ units [eq.~(\protect\ref{eq:efficiency})] is shown
  as blue triangles, solid connecting lines on the left vertical axis.
  Maximum $\zeta$ efficiency of $N$ units
  [eq.~(\protect\ref{eq:efficiency-2})] for $P_c=0.5$ (red circles,
  dashed) and $P_c=0.3$ (red squares, dotted) on the right vertical
  axis.}
\label{fig:efficiency}
\end{figure}

On the other hand, what if we do know the Lagrangian characteristics
of the flow?  Now we can choose the most efficient combination of $N$
sites, i.e.\ those that remove the most pollutant after a long time, or
those that reach the target $P_c$ the fastest.  There are several ways
to define efficiency.  We can define the efficiency of a set of $N$
units as the total amount captured divided by $N$, i.e.\
\begin{equation}
\label{eq:efficiency}  
  \eta(m_1,\dots,m_N) = \frac{1-P_\infty}{N},
\end{equation}
which is the average removal per unit.  Fig.~\ref{fig:efficiency}
shows that in all three cases (non-chaotic, mixed, and chaotic), as $N$ increases, the maximum $\eta$ efficiency (blue
triangles, left vertical axis) decreases.  This is because the regions
cleaned by different units overlap, and there is generally more
overlap when more units are used.  In the facility location problem
this ``diminishing returns'' property is called submodularity.  For
general flows with chaotic and regular regions, overlapping capture
regions will be unavoidable, even when the best combination of unit
locations is chosen.

We can also define the efficiency of a set of $N$ units as the rate at
which they approach success (reach $P_c$) by
\begin{equation} \label{eq:efficiency-2}
\zeta(m_1,\dots,m_N) = \frac{1-P_c}{N\tau},
\end{equation}
where $\tau$ is the time required to reach $P_c$.
Eq.~(\ref{eq:efficiency-2}) measures rate efficiency of the capture
process, whereas eq.~(\ref{eq:efficiency}) measures total long-time
efficiency.  Unlike the maximum $\eta$ efficiency, the maximum $\zeta$ efficiency does not suffer from diminishing returns, as shown by the red circles and squares in Fig.~\ref{fig:efficiency}.  In all cases shown in Fig.~\ref{fig:efficiency}, the maximum $\zeta$ efficiency initially increases with $N$.  This means that using more units is more efficient for
short-term goals, but in the long-term fewer units is more efficient.  For example, an increase in $\zeta$ from $N=1$ to $N=2$, as occurs in the mixed and chaotic cases [Fig.~\ref{fig:efficiency}(b,c)], means the rate of capture more than doubles when the number of units is doubled.  However, in the mixed case with $P_c=0.5$, and the two chaotic cases, $\zeta$ does not increase monotonically.  This suggests that there is an optimal choice of $N$ at which peak $\zeta$ efficiency is achieved, e.g.\ $N=4$ in the mixed case with $P_c=0.5$.

%Maximum $\zeta$ efficiency increases with $N$ for both
%$P_c=0.5$ and $P_c=0.3$, as shown by the red circles and squares in
%Fig.~\ref{fig:efficiency}.  Using more units is more efficient for
%short-term goals, but in the long-term fewer units is more efficient.
%Efficiency varies with the required capture rate, and capture should
%begin as early as feasible, in order to take advantage of the
%long-term efficiency.

\subsection{Heuristic methods to find near-optimal site locations}

Another important question is how to choose the most effective $N$
sites based \textit{only} on characteristics of the flow, rather than
measuring $P_\infty$ for all possible site combinations?  At short
times, pollutant captured by a unit is proportional to the total volume of
fluid that passes through it.  Therefore, best short term results are
obtained by locating units in regions with high velocity.  However, at
long times, the total volume of fluid that passes through a unit is
not a good predictor of performance, because clean fluid may
continually recycle through units.  

%For example, the four units
%closest to the origin in Fig.~\ref{fig:colored_machine_locations}
%achieve $P_\infty = 0.55$, i.e.\/ they remove $45\%$ of the CO$_2$
%from the domain.  Exactly the same volume of fluid flows through the
%units closest to the points $(0,\pm 1)$ and $(\pm 1,0)$; however, due
%to the constant recirculation of clean fluid, they only achieve
%$P_\infty = 0.90$, meaning they only remove $10\%$ of the CO$_2$.
%Recirculation is bad, and units should not be located where it occurs. {\bf ::Note that in flows where is mostly recirculation, and little chaos, it is necessary, and may be preferable, to capture recirculating regions.::}

\begin{figure}[tb]
\centering
\includegraphics[width=0.45\textwidth]{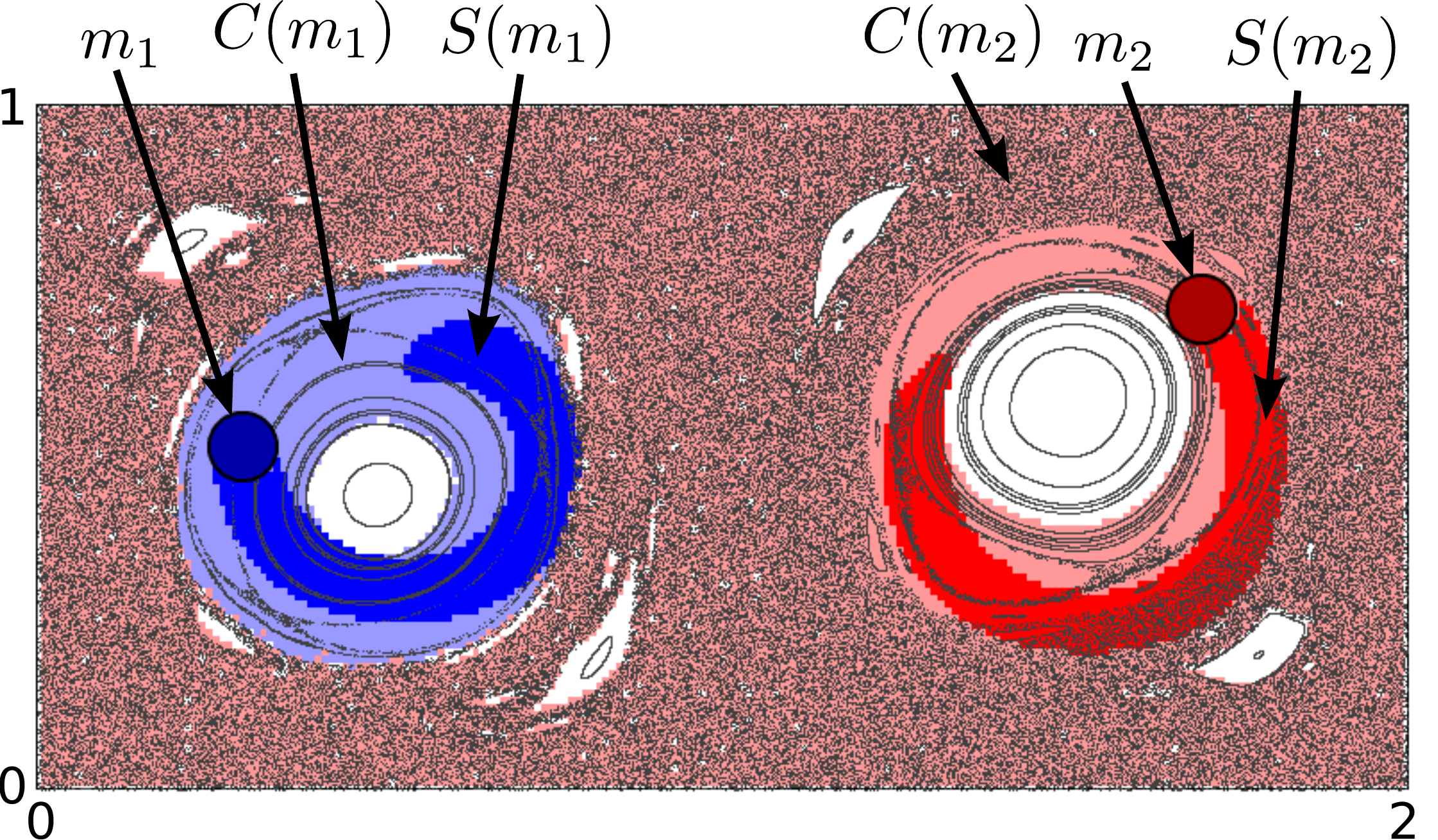}
\caption{The captured region $C(m)$ (light blue and light red) for two
  intake units, $m=m_1$ (blue circle) and $m=m_2$ (red circle).  $C(m)$ is
  the infinite-time streak-surface emanating from the unit, which is
  equivalent to the union of iterates of $S(m)$
  [eq.~(\protect\ref{eq:cleaned_region})], the set of streak
  trajectories emanating from the unit over one flow period (dark blue
  and dark red).  For $m_1$, $S(m_1)$
  intersects coherent regions in the left island, so $C(m_1)$ contains a portion of the left island. However, $S(m_1)$ does not intersect the chaotic region, so it
  is not captured.  For $m_2$, $S(m_2)$ intersects coherent regions in the right island: $C(m_2)$ contains a portion of the right island.  As $S(m_2)$ also intersects a portion of the chaotic
  region, the entire chaotic region is contained in $C(m_2)$, even
  though $m_2$ is inside the right island.}
\label{fig:cleaned_region}
\end{figure}

For time-periodic flows, such as the double gyre example considered here,
we can predict the long term performance of any proposed site using
flow properties.  The captured region is the infinite-time
streak-surface that emanates from a unit $m$, which can be written as the
set
\begin{equation}  \label{eq:cleaned_region}
C(m)=\bigcup_{n=0}^\infty T^n S(m)
\end{equation}
where $T$ is the map that takes a particle to its position after one
flow period, and $S(m)$ is the set of streak trajectories emanating
from a unit over one flow period: $C(m)$ is the union of all chaotic
and coherent regions (invariant tori) that intersect $S(m)$. Importantly, $C(m)$
contains the entire chaotic region if and only if $S(m)$ intersects
some portion of the chaotic region \footnote{Technically, $C(m)$
  contains the entire chaotic region except a set of measure zero,
  known as the unstable manifold of the chaotic saddle, and the area
  of the captured region converges asymptotically as $n\to \infty$
  \cite{Altmann2013leaking}.}.  Two examples are shown in
Fig.~\ref{fig:cleaned_region}.  For the
unit $m_1$ (dark blue circle), $S(m_1)$ (dark blue) intersects invariant tori in the left island, so
$C(m_1)$ (light blue) contains all those invariant tori, i.e.\ it contains a portion of the left island.  However,
$S(m_1)$ does not intersect the chaotic region, so it is not captured.  For the capture unit $m_2$ (dark red
circle), $S(m_2)$ (dark red) intersects some invariant tori in the right island, so $C(m_2)$ (light red)
contains all those invariant tori, i.e.\ it contains a portion of the right island.  $S(m_2)$ also intersects the
chaotic region, so $C(m_2)$ contains the entire chaotic region, even
though $m_2$ itself is contained within the right island.  There are also coherent regions uncaptured by either unit.

Based on eq.~(\ref{eq:cleaned_region}), a good individual
site is one for which $S(m)$ intersects a portion of the chaotic
region and a collection of coherent regions with a large area.  For $N\geq 2$, the best combination of sites is not the same as
choosing the best $N$ individual sites, as they may capture the same
region.  For increasing $N$, predicting the best sites becomes an
increasingly difficult optimization problem.  One heuristic approach, which uses the Poincar\'{e} section and eq.~(\ref{eq:cleaned_region}), is to choose a first site $M_1$ such that $S(M_1)$ intersects a portion of the chaotic
region and the collection of coherent regions with the largest area.  The site $M_1$ captures the entire chaotic region and some coherent regions. For example, the site $m_2$ in Fig.~\ref{fig:cleaned_region} is a good candidate.  Next, choose a site $M_2$ such that $S(M_2)$ intersects coherent regions with the largest area, excluding the coherent regions that are intersected by $S(M_1)$.  For example, the site $m_1$ in Fig.~\ref{fig:cleaned_region} is a good candidate for the second site because $S(m_1)$ intersects the outer edge of the left island, which is not captured by $m_2$.  Continue choosing sites $M_i$ such that $S(M_i)$ intersects coherent regions with largest area that are not captured by the sites $M_1,\dots,M_{i-1}$.  This procedure is summarized in Algorithm~\ref{alg:find_optimum_guess}.  Note that a computationally simpler approach is to replace the sets $S(M_i)$ in the algorithm with the sets $M_i$, since $M_i$ is necessarily contained in $S(M_i)$.  This alternative approach requires only the Poincar\'{e} section to be computed, and not the sets $S(M_i)$.  However, the locations chosen will perform worse than if the sets $S(M_i)$ are used.

\begin{algorithm}[H]
\caption{Choose $N$ sites based on qualitative properties of the Poincar\'{e} section and eq.~(\ref{eq:cleaned_region})}
\label{alg:find_optimum_guess}
\begin{algorithmic}[1]
\State{Generate the Poincar\'{e} section}
\State{Choose the first site, $M_1$ such that $S(M_1)$ intersects the chaotic region and invariant tori whose union has the largest area}
\For{$i=2,\dots,N$}
\State{Choose site location $M_i$ such that $S(M_i)$ intersects invariant tori whose union has the largest area, excluding invariant tori intersected by $S(M_1),\dots,S(M_{i-1})$}
\EndFor
\State \Return Sites $M_1,\dots,M_N$
\end{algorithmic}
\end{algorithm}

\begin{figure}[tbp]
\centering
\includegraphics[width=\columnwidth]{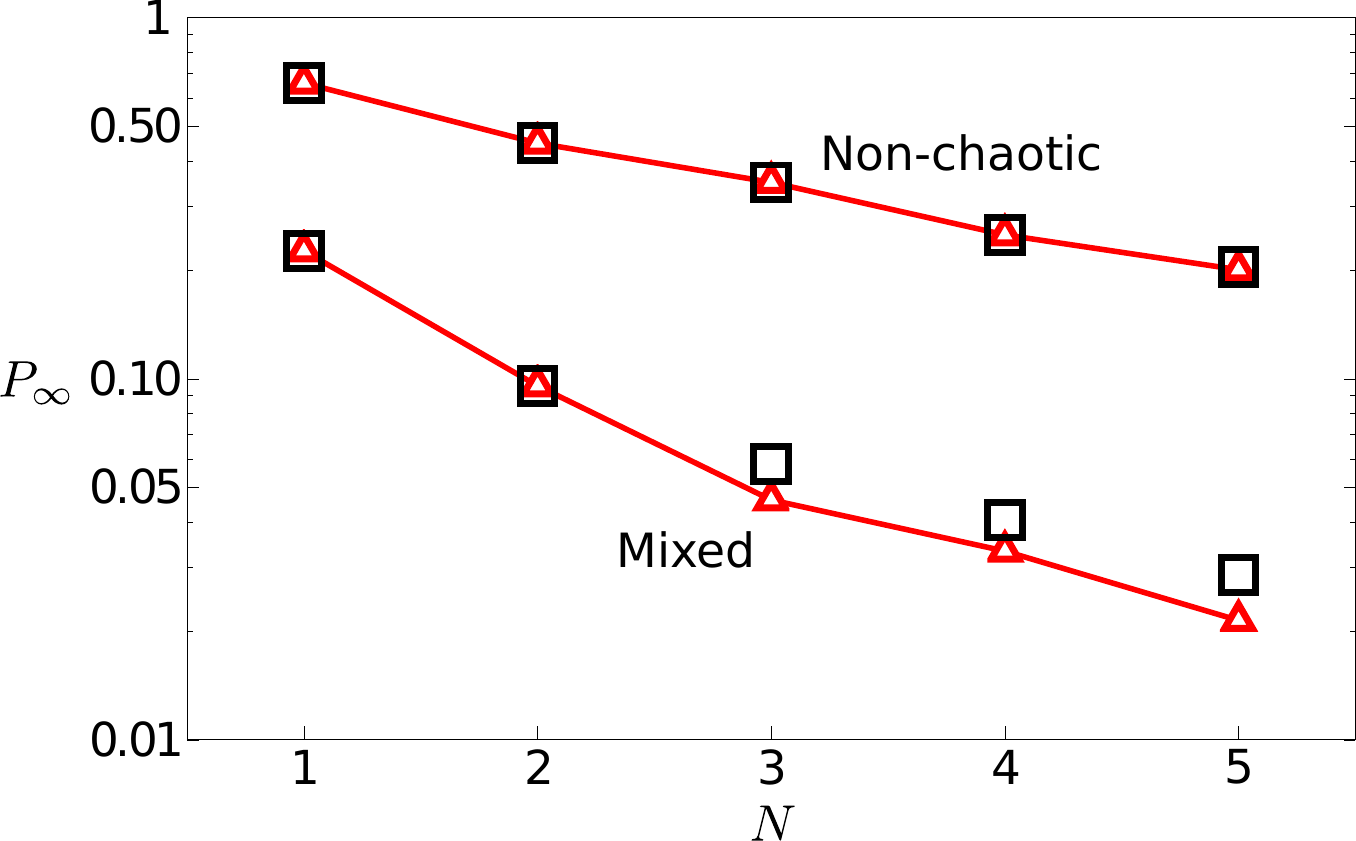}
\caption{Minimum $P_\infty$ over all combinations of $N$ units (red triangles), and $P_\infty$ obtained from the heuristic approach described in Algorithm~\ref{alg:find_optimum_iterative} (black squares).  Shown for the non-chaotic (upper) and mixed (lower) cases.}.
\label{fig:global_optima_vs_heuristic}
\end{figure}

Consider a second heuristic approach. First, choose the best individual site, $M_1$, that
minimizes $P_\infty$. Equivalently, $M_1$ maximizes the captured area, $C(M_1)$. Then choose a next site, $M_2$, such that the pair maximizes the captured area, $C(M_1) \cup C(M_2)$.  Continue choosing sites $M_i$ that maximize the captured area, $\bigcup_{j=1}^i C(M_j)$, until $N$ or the capture target, $P_c$, is reached.  This procedure is summarized in Algorithm~\ref{alg:find_optimum_iterative}.  Due to submodularity of $\eta$, the amount of
pollutant removed using this algorithm will always be within a constant
factor of the optimal solution \cite{Fujishige_submodular_2005}.  Fig.~\ref{fig:global_optima_vs_heuristic} shows the effectiveness of Algorithm~\ref{alg:find_optimum_iterative} for the non-chaotic and mixed cases.  The global optima for $P_\infty$ (shown as red triangles), found by minimizing $P_\infty$ over all combinations of $N$ locations, are only slightly less than the values obtained from the heuristic approach (shown as black squares).  Comparing the computational expense of the two methods, for the global optima, the number of combinations with $J=50$ possible site locations grows as the binomial coefficient, whereas the heuristic approach requires only $JN$ evaluations.  For $N=5$, this is the difference between over two million evaluations versus $250$ evaluations.

\begin{algorithm}[H]
\caption{Choose $N$ sites by computing capture regions}
\label{alg:find_optimum_iterative}
\begin{algorithmic}[1]
\State{Choose $J$ potential site locations, $M_1,\dots,M_J$, e.g.\ on a grid}
\State{Compute the captured region, $C(M_i)$, for each site using particle tracking or eq.~(\ref{eq:cleaned_region})}
\For{$i=1,\dots,N$}
\State{Choose site location $M_i$ such that $\bigcup_{j=1}^i C(M_i)$ has the largest area.}
\EndFor
\State \Return Sites $M_1,\dots,M_N$
\end{algorithmic}
\end{algorithm}

The heuristic algorithms described above do not account for mixing
rate, and optimizing $\zeta$ is in general more difficult because it
is not submodular (it does not decrease monotonically with $N$).  More
sophisticated mathematical techniques could be useful for different
applications of the capture problem.

\section{Conclusions}

Capturing pollutants from a heterogenous flow environment is far from
simple.  As our example problem is highly stylized (perfect capture
units in a relatively simple time-periodic flow), the absolute numbers
from our calculations are not the main point.  The important features
of the flow capture problem illustrated here are the trends that we
expect to persist in more realistic models and across various
applications.  For example, we see the same trends and qualitative
results when the double gyre flow is replaced with the ``egg-beater''
flow \cite{Franjione1992}, which uses periodic boundary conditions
(see Supplemental Material when published).
The similarities between our results for the double gyre flow and the
egg-beater flow show that the geometry of transport structures (e.g.\
islands and the chaotic sea) plays a more important role for the flow
capture problem than the specifics (e.g.\ boundary conditions) of the
flows that generate them.

When flow heterogeneity is accounted for, the maximum capture
efficiency depends on the required capture rate.  Fast capture rates
require more units, but over long times, fewer units are more
efficient.  If efficiency---often related to cost---is important, the
removal process should begin as early as feasible.  

Dynamical systems (or Lagrangian coherent structures in aperiodic flows) offers a path
to predictability for capture unit placement because streak surfaces
(manifolds) shape transport in flows \cite{Wiggins2}.  We have demonstrated two heuristic algorithms to near-optimally position capture units. Both algorithms significantly reduce the computational cost of optimization compared to checking all possible combinations of locations.  Most
importantly, for flows with transport barriers (e.g.\ islands), placing units in random locations without accounting for
the heterogeneity and chaos in the flow carries a remarkably high risk
of failure.  Even when units capture perfectly, as we assumed, or they
receive large flow volumes, they may not be effective without
accounting for local heterogeneity and chaos in placement, design, and
operation.

For fully chaotic flows, single capture units positioned arbitrarily
will eventually capture \emph{all} the pollutant.  Hence, for
long-term capture, there is no benefit in optimizing unit location.
However, the rate of capture depends on unit location, so some
locations are better than others for reaching goals rapidly.  We have
not explored this dependence in detail here, and it is an important
direction for future study.  In addition, since every capture unit
location will eventually capture all the pollutant, it would seem that
there is no benefit to using multiple capture units.  However, we have
shown that using two units can more than double the rate of capture.
Therefore, there can be benefits to using multiple capture units,
especially if capture must occur within a specified time.

Future work should explore more complex models, with more realistic
velocity fields derived for specific applications, such as
microplastic capture in the ocean \cite{Andrady2011microplastics,
  Cozar2014microplastics} and direct capture of CO$_2$ from the
atmosphere \cite{Socolow2011direct, IPCC_2014, Smith_limits_2015,
  Keith2018}. In addition, the effects of turbulent diffusion
\cite{Lester2009, Schlick2013}, heterogeneous initial pollutant
distributions, or pollutant sources in addition to sinks
\cite{Thiffeault2008} should also be considered.  Furthermore, flow
capture applications where the pollutant is not a passive tracer pose
additional challenges, e.g.\ capturing inertial microplastic particles
\cite{Andrady2011microplastics, Cozar2014microplastics} or motile
particles \cite{Pedley1992motile, Torney2007motile,
  Khurana2011motile}.  Insights into the general flow capture problem
could also be gained by using quantitative Poincar\'{e} recurrence
theory \cite{Saussol2009} to analyse removal rates and recirculation
rates of capture units in heterogeneous flows.

%\bibliographystyle{unsrt}
%\bibliographystyle{ieeetr}
%\bibliography{mybib2}

\end{document}